\documentclass[12pt,showkeys,showpacs,indentfirst,bm]{revtex4}
\usepackage{epsfig}
\usepackage{epsf}
\usepackage{bm}
\topmargin=3.0cm \Roman{table}

\newcommand{\al}{&\!\!\!\!~}
\newcommand{\be}{\begin{equation}}
\newcommand{\ee}{\end{equation}}
\newcommand{\ba}{\begin{eqnarray}}
\newcommand{\ea}{\end{eqnarray}}
\newcommand{\noo}{\nonumber}

\begin{document}

\title{$S$ wave $K\pi$ scattering and effects of $\kappa$ in $J/\psi\to {\bar
K}^{*0}(892)K^+\pi^-$}

\author{F.-K. Guo$^{1,2,6}$}
\email{guofk@mail.ihep.ac.cn}
\author{R.-G. Ping$^{1,2}$}
\author{P.-N. Shen$^{1,2,4,5}$}
\author{H.-C. Chiang$^{3,2}$}
\author{B.-S. Zou$^{1,2}$}
\affiliation{\small
$^1$CCAST(World Lab.), P.O.Box 8730, Beijing 100080, China\\
$^2$Institute of High Energy Physics, Chinese Academy of Sciences,
P.O.Box 918(4), Beijing 100049, China\footnote{Corresponding address.}\\
$^3$South-west University, Chongqing 400715, China\\
$^4$Institute of Theoretical Physics, Chinese Academy of Sciences, P.O.Box 2735, China\\
$^5$Center of Theoretical Nuclear Physics, National Laboratory of
Heavy Ion Accelerator, Lanzhou 730000, China\\
$^6$Graduate University of Chinese Academy of Sciences, Beijing
100049, China}
\date{\today}

\vspace{1cm}

\begin{abstract}
$K\pi$ $S$ wave scattering is studied using a chiral unitary
approach (ChUT) taking into account coupled channels. With the
amplitudes derived from the lowest order chiral Lagrangian as the
kernel of a set of coupled channel Bathe-Salpeter equations, the
$I=1/2$ $S$ wave $K\pi$ scattering phase shifts below $1.2$ GeV
can be fitted by one parameter, a subtraction constant, and a
scalar resonance corresponding to the controversial $\kappa$
($K^*_0(800)$) can be generated dynamically. A good description of
the $I=3/2$ $S$ wave $K\pi$ scattering phase shifts below $1.2$
GeV can also be obtained. An artificial singularity in the
conventional cut-off method of the 2-meson loop integral of the
ChUT is found. The formalism is applied to deal with the $S$ wave
$K\pi$ final state interaction (FSI) in the decay $J/\psi\to {\bar
K}^{*0}(892)K^+\pi^-$, and a qualitatively good fit to the data is
achieved. The role of $\kappa$ in the decay is discussed.
\end{abstract}


\maketitle

\section{Introduction}

The scattering between the lowest pseudoscalar mesons is
significant in the understanding of low energy strong interaction
and nonperturbative QCD. The most popular approach dealing with
this problem is Chiral Perturbation Theory
(ChPT)\cite{we79,gl84,gl85,me93,pi95} (for a recent comprehensive
review, see \cite{sche}). The lowest pseudoscalar meson octet
($\pi,~K,~\eta$) is identified with the Goldstone bosons
associated with the spontaneous breakdown of chiral symmetry of
QCD. The most general chiral Lagrangian can be written in a
perturbative manner to an order by expanding the QCD Lagrangian
containing external sources in powers of the external momenta of
the pseudoscalar mesons and of the light quark masses $m_u$, $m_d$
and $m_s$. However, a perturbative expansion to any finite order
can not describe the appearance of a resonance. For instance, the
pure ChPT is not suitable for the physics of the scalar isoscalar
$\pi\pi$ scattering above $0.6$ GeV where a broad resonance
$f_0(600)$ (or $\sigma$) is generally believed to exist.
Furthermore, unitary relation of $S$ matrix is only respected by
ChPT in a perturbative sense. Over the last few years, several
nonperturbative methods were proposed to extend the chiral
expansion to higher energies, such as resummation of loops using a
Bethe-Salpeter equation (BSE) \cite{oo97,na00}, the inverse
amplitudes method (IAM) \cite{tr88,oo99,go99,np02} and dispersion
relation methods \cite{oond,xz01,zz04}.

The existence of the scalar particle $\kappa$ (i.e. $K^*_0(800)$)
and the mass and width of this particle, if it exists, has been
long controversial since the 1970's \cite{br86,tr96,br99}. The
evidences of the $\kappa$ has been observed in the analysis of the
$K\pi$ scattering phase shifts \cite{lass}, the Dalitz Plot
Analysis of the Decay $D^+\to K^-\pi^+\pi^+$ \cite{e791}, and the
BES data of the $J/\psi$ decays \cite{wu03,bugg,bes05}.

In order to understand the $\kappa$, a good understanding of $S$
wave $K\pi$ scattering is required. $K\pi$ scattering has been
investigated using the ChPT by V. Bernard {\it et al.} at
next-to-leading order \cite{bkm} and a satisfactory description of
the available data up to about 900 MeV was found, and by J. Bijnens
{\it et al.} at next-to-next-to-leading order \cite{bdt}. The
problem has also been investigated using various unitarisation
approaches of the chiral Lagangian \cite{oo99,oond,jop00,np02,zz04},
and by other methods such as the interfering Breit-Wigner amplitude
method \cite{ish}, the unitarized meson model \cite{br01} and the
$K$-matrix method \cite{lzl03}. A scalar meson corresponding to
$\kappa$ was found in all of the mentioned unitary chiral
approaches. While Long Li {\it et al.} found the $K\pi$ data could
be fitted without a pole around 900 MeV in their $K$-matrix approach
and only one s-channel resonance between the $K\pi$ threshold and
1.6 GeV, i.e., $K^{0*}(1430)$ with a mass around 1438-1486 MeV and a
width of about 346 MeV was found \cite{lzl03}. Recently, an
important work appeared. Some authors have performed a detailed
analysis using a set of Roy-Steiner (RS) equations \cite{bd04}. The
input was high-statistical data at $\sqrt{s}\gtrsim1$ GeV for both
$\pi K\to\pi K$ and $\pi\pi\to K{\bar K}$ amplitudes, then the
amplitudes below 1 GeV were determined. The authors computed $S$
wave scattering lengths and the coupling constants $L_1$, $L_2$,
$L_3$ and $L_4$ appearing in the $O(p^4)$ order chiral Lagrangian.
The precise $S$ wave and $P$ wave phase shifts below about 1 GeV
were also obtained by solving the RS equations.

In this paper, we use the on-shell coupled-channel BSE approach
(it will be referred as ChUT below) proposed by J. A. Oller and E.
Oset \cite{oo97} to study $S$ wave $K\pi$ scattering. In
Ref.~\cite{oo97}, the lowest order ChPT amplitudes are employed as
the kernels of a set of coupled channel on-shell BSEs. The BSEs
are solved to resum the contributions from the s-channel loops of
the re-scattering between pseudoscalar mesons. The loop integral
was calculated using a three-momentum cut-off, and this is the
only free parameter in the approach. By properly choosing the
cut-off, the ChUT can well-describe the data of the $S$ wave
meson-meson interaction up to $\sqrt{s}\simeq1.2$ GeV in the
isoscalar and isovector channels. And more interestingly, the
$\sigma$, $f_0(980)$ and $a_0(980)$ scalar mesons can be generated
dynamically as the poles in the second Riemann sheet of the $T$
matrix. It was mentioned in Ref.~\cite{lo04} that satisfactory
fits to the $K\pi$ scattering phase shifts could be produced using
a cut-off of $850$ MeV and using $f=100$ MeV (an average between
$f_{\pi}$ and $f_K$). However, we find there is a severe problem
in the cut-off method in calculating the loop integral, which makes
it difficult to describe $K\pi$ data above 1 GeV, and we will discuss
this in detail in the corresponding section.

On the other hand, $J/\psi$ radiative and hadronic decays provide
a rather rich source of the information of light hadrons including
$\sigma$ and $\kappa$, and hence a good understanding of the
interaction between the final state hadrons is required. The ChUT
has been used to dealing with the meson-meson $S$ wave FSI in
$\gamma\gamma\to$ meson-meson reactions \cite{oo98}, in some heavy
quarkonium decays \cite{mo01,lov04,rp04,gscp}, and in heavy meson
decays \cite{gm02,ol05}.

The paper is organized as follows: The basic definitions and the
lowest order chiral amplitudes to be used are given in
Sec.~\ref{sec:amp}. In Sec.~\ref{sec:chut}, we give a simple
review of the ChUT first, and then in Sec.~\ref{sec:com} we
discuss an artificial singularity appeared in the conventional
cut-off method used in the ChUT, and a very simple proof of the
unitarity is given in Sec.~\ref{sec:proof}. The calculated $S$
wave $K\pi$ phase shifts are shown in Sec.~\ref{sec:ps}
and~\ref{sec:ps2}, and a good description of the experimental data
below 1.2 GeV is achieved. The pole position associated to
$\kappa$ is given in Sec.~\ref{sec:pole}. As an example of
applying the ChUT formalism to dealing with the $S$ wave $K\pi$
FSI, the decay $J/\psi\to {\bar K}^{*0}(892)K^+\pi^-$ is discussed
in Sec.~\ref{sec:app}. A brief summary is given in
Sec.~\ref{sec:con}.

\section{\label{sec:amp}The lowest order ChPT amplitudes}

\subsection{Definitions}

The lowest order, i.e. $O(p^2)$, Lagrangian for SU(3) ChPT reads
\be%
\label{eq:LU}
{\cal L}^{(2)}=\frac{f^2}{4}\langle\partial_{\mu}U\partial^{\mu}U^{\dag}+M(U+U^{\dag})\rangle,%
\ee%
where $f$ is the pion decay constant, the physical value of which
is 92.4 MeV, and $\langle\rangle$ stands for the trace of
matrices. The $3\times 3$ special unitary matrix $U$ is defined as
$U=\exp{\frac{i\sqrt{2}}{f}\Phi}$, and $\Phi$ is made of the
pseudo Goldstone boson fields
\be%
\label{eq:ps} \Phi = \left(
\begin{array}{ccc}
\frac{1}{\sqrt{2}}\pi^0 + \frac{1}{\sqrt{6}}\eta_8 & \pi^+ & K^+\\
\pi^- & -\frac{1}{\sqrt{2}}\pi^0 + \frac{1}{\sqrt{6}}\eta_8 & K^0\\
K^- & \bar{K}^0 & - \frac{2}{\sqrt{6}}\eta_8
\end{array}
\right).
\ee%
The pseudoscalar meson mass matrix $M$ is given by
\be%
\label{eq:phi}
M = \left(
\begin{array}{ccc}
m_{\pi}^2 & 0 & 0\\
0 & m_{\pi}^2 & 0\\
0 & 0 & 2m_K^2-m_{\pi}^2
\end{array}
\right),%
\ee%
where the isospin limit $m_u=m_d$ is assumed. The relevant part
concerns four mesons and can be derived from Eq.~(\ref{eq:LU}) as
\be%
\label{eq:L2} {\cal
L}^{(2)}=\frac{1}{12f^2}\langle(\partial_{\mu}\Phi\Phi -
\Phi\partial_{\mu}\Phi)^2 + M\Phi^4\rangle.
\ee%

Let $T^I(s,t,u)$ denotes the amplitude with total isospin $I$ and
$T^I_l(s)$ its partial wave projection with angular momentum $l$ .
We take the normalization for partial wave amplitudes in such a
way that
\be%
T^I_l(s)=\frac{1}{2}\int^{1}_{-1}d\cos{\theta}P_l(\cos{\theta})
T^I(s,t(s,\cos{\theta}),u(s,\cos{\theta})),%
\ee%
where $P_l(\cos{\theta})$ is the $l$-th order Legendre polynomial,
$\theta$ is the scattering angle in the center of mass frame. The
Mandelstam variables $s=(p_1+p_2)^2$, $t=(p_1-p_3)^2$ and
$u=(p_1-p_4)^2$ are bound to the identity
$s+t+u=\sum_{i=1}^{4}m_i^2$ when the particles are on-shell, and
\ba%
t(s,\cos{\theta})=\al\al m_1^2+m_3^2 -
2\sqrt{[m_1^2+\frac{\lambda(s,m_1^2,m_2^2)}{4s}]
[m_3^2+\frac{\lambda(s,m_3^2,m_4^2)}{4s}]} \noo\\
\al\al+\frac{1}{2s}\sqrt{\lambda(s,m_1^2,m_2^2)\lambda(s,m_3^2,m_4^2)}\cos{\theta}
\ea%
with $\lambda(s,m_i^2,m_j^2)=[s-(m_i+m_j)^2][s-(m_i-m_j)^2]$.

We shall use the phase convention $|\pi^+\rangle=-|1,1\rangle$ and
$|{\bar K}^0\rangle=-|1/2,1/2\rangle$. The calculations of $I=1/2$
scattering will be performed with single channel ($K\pi$), two
channels ($K\pi$ and $K\eta$) and three channels ($K\pi$, $K\eta$
and $K\eta'$), respectively, and the single channel $I=3/2$ $K\pi$
scattering will also be investigated. The lowest order chiral
amplitudes to be used in our coupled channel calculations of the $S$
wave $K\pi$ scattering will be given in the following subsections.
We will label them with $V$ and label the full amplitudes in the
BSEs with $T$ to avoid confusion. Since we will restrict ourselves
in the $S$ wave, the subscript $l$ will be neglected in the
following and $l=0$ will be implied everywhere.

\subsection{Two channels case}

In this case, we take $\eta=\eta_8$ and the mixing between $\eta$
and $\eta'$ will not be considered. Let us label the channels
$K\pi$ and $K\eta$ with 1 and 2 respectively, such that
$V_{11}^I(s,t,u)$, $V_{12}^I(s,t,u)$ and $V_{22}^I(s,t,u)$
represent the amplitudes with total isospin $I$ for $K\eta\to
K\eta$, $K\eta\to K\pi$ and $K\eta\to K\eta$. By time reversal
invariance, one has $V_{ij}^I(s,t,u)=V_{ji}^I(s,t,u)$.

$\pi^+K^+\to\pi^+K^+$ is a pure $I=3/2$ process,
\be%
\label{eq:piK}
V^{3/2}(s,t,u)=V_{\pi^+K^+\to\pi^+K^+}(s,t,u).%
\ee%
According to the crossing symmetry, the isospin relation for
$V^{1/2}_{11}(s,t,u)$ and $V^{1/2}_{12}(s,t,u)$ can be interpreted
as
\ba%
V^{1/2}_{11}(s,t,u) &=& \frac{3}{2}V^{3/2}(u,t,s)-\frac{1}{2}V^{3/2}(s,t,u),\\
V^{1/2}_{12}(s,t,u) &=& \sqrt{3}V_{K^0\eta\to K^0\pi^0}(s,t,u).
\ea%

Now the $O(p^2)$ order ChPT amplitudes can be derived from
Eq.~(\ref{eq:L2}). One has a pure $S$ wave amplitude for $I=3/2$
process
\be%
V^{3/2}(s,t,u)=\frac{1}{2f^2}(s-m^2_{\pi}-m^2_K),
\ee%
and for $I=1/2$ the amplitudes contain both $S$ wave and $P$ wave
\ba%
\al\al V^{1/2}_{11}(s,t,u)=-\frac{1}{4f^2}(4s+3t-4m^2_{\pi}-4m^2_K),\label{eq:v11} \\
\al\al V^{1/2}_{12}(s,t,u) =
-\frac{1}{4f^2}(-3t+\frac{1}{3}m^2_{\pi}+\frac{8}{3}m^2_K+m^2_{\eta}),\label{eq:v12}\\
\al\al V^{1/2}_{22}(s,t,u) =
-\frac{1}{4f^2}(3t-\frac{2}{3}m^2_{\pi}-2m^2_{\eta}).
\ea%
Note that the calculations will be performed using on-shell BSE,
so that the on-shell condition $p_i^2=m_i^2$ has been used in
writing the above amplitudes. The isospin phase convention of $K$
is different from the one in Ref.~\cite{oo99}, such that there is
an extra minus before Eq.~(\ref{eq:v12}) compared with
$T_2(s,t,u)$ in Eq.~(B14) therein, but the physical observables
are not influenced by different phase conventions as they should
be.

\subsection{Three channels case}

If one want to see the possible impacts of the $K\eta'$ channel,
$\eta'$ should be included in the chiral Lagrangian. One way is
extending the $SU(3)$ matrix to a $U(3)$ one
\ba%
\Phi_{nonet} \al\al= \Phi + \frac{I}{\sqrt{3}}\eta_0 \noo\\
\al\al=
\left(
\begin{array}{ccc}
\frac{1}{\sqrt{2}}\pi^0 + \frac{1}{\sqrt{3}}\eta + \frac{1}{\sqrt{6}}\eta' & \pi^+ & K^+\\
\pi^- & -\frac{1}{\sqrt{2}}\pi^0 + \frac{1}{\sqrt{3}}\eta + \frac{1}{\sqrt{6}}\eta' & K^0\\
K^- & \bar{K}^0 & -\frac{1}{\sqrt{3}}\eta +
\frac{2}{\sqrt{6}}\eta'
\end{array}
\right),
\ea%
where the standard $\eta-\eta'$ mixing
\ba%
|\eta\rangle\al=\al\frac{1}{3}|\eta_0\rangle+\frac{2\sqrt{2}}{3}|\eta_8\rangle,\noo\\
|\eta'\rangle\al=\al\frac{2\sqrt{2}}{3}|\eta_0\rangle-\frac{1}{3}|\eta_8\rangle
\ea%
is assumed. We label the channel $K\eta'$ with 3, such that
$V_{13}^{1/2}(s,t,u)$, $V_{23}^{1/2}(s,t,u)$ and
$V_{33}^{1/2}(s,t,u)$ represent the amplitudes for $K\eta'\to
K\pi$, $K\eta'\to K\eta$ and $K\eta'\to K\eta'$, respectively. The
amplitudes concerning only $K$ and $\pi$ do not change while the
others change to
\ba%
\al\al V^{1/2}_{12}(s,t,u) =
-\frac{\sqrt{2}}{6f^2}(-3t+2m^2_K+m^2_{\eta}),\\
\al\al V^{1/2}_{22}(s,t,u) =
-\frac{2}{9f^2}(3t-m^2_K-2m^2_{\eta}).
\ea%
The amplitudes involving $\eta'$ are
\ba%
\al\al V_{13}^{1/2}(s,t,u)=
\frac{1}{12f^2}(-3t+3m^2_{\pi}+8m^2_K+m^2_{\eta'}),\\
\al\al V_{23}^{1/2}(s,t,u)= \frac{\sqrt{2}}{18f^2}(3t - 3m^2_{\pi}
+2m_K^2-m^2_{\eta}-m^2_{\eta'}),\\
\al\al V_{33}^{1/2}(s,t,u) = -\frac{1}{12f^2}(t-2m^2_{\pi} +
12m_K^2+\frac{2}{3}m^2_{\eta'}).
\ea%

\section{\label{sec:chut}Coupled-channel chiral unitary approach}

In our normalization, the unitary relation for the partial
amplitudes with isospin $I$ satisfies
\ba%
\label{eq:ur}
Im T^I(s)=-T^{I\dagger}(s)\rho(s)T^I(s),%
\ea%
where $T^I(s)$ and $\rho(s)$ are $n\times n$ matrices for $n$
channel calculations. The matrix elements $T^I_{ij}(s)$ will be
given by the on-shell BSEs. $\rho(s)$ is a diagonal matrix
with%
\be%
\rho_{ii}(s)=\frac{{p_{cm}}_i}{8\pi
\sqrt{s}},%
\ee%
where ${p_{cm}}_i$ is the three-momentum of one meson in the
center-of-mass (c.m.) frame for the $i$-th channel. This momentum
is given by
\be%
p_{cm}=\frac{1}{2\sqrt{s}}\sqrt{[s-(m_1+m_2)^2][s-(m_1-m_2)^2]}
\ee%
in a channel with two mesons of masses $m_1$ and $m_2$
respectively. In what follows we will omit the label $I$ for
simplication.

Following Ref.~\cite{oo97}, the amplitude of meson-meson
scattering can be cast using the BSE
\be%
\label{eq:bse}
T=(1-VG)^{-1}V,%
\ee%
where $V$ is the lowest order chiral amplitude and $G$ is the loop
propagator. For coupled channel calculations, all of $T$, $V$ and
$G$ are matrices. One important feature of this approach is that
$V$ can be factorized on shell in the BSEs, and so that the
integral equations become algebraic equations which can be solved
simply. This feature is justified using the $N/D$ method of
dispersion relations \cite{oond} and by a comprehensive treatment
of the BSE \cite{na00}.

\subsection{\label{sec:com}Artificial singularity in the cut-off method}

The loop integral in the $i$-th channel is
\begin{equation}
G_{ii}(s)=i\int\frac{d^4q}{(2\pi)^4}\frac{1}{q^2-m_1^2+i \varepsilon} \frac{1}{%
(p_1+p_2-q)^2-m_2^2+i\varepsilon},  \label{eq:2loop}
\end{equation}
where $p_1$ and $p_2$ are the four-momentum of the two initial
particles respectively, $m_1$ and $m_2$ are the masses of the two
particles appearing in the loop. Sometimes in ChUT the integral is
calculated by using a three-momentum cut-off parameter $q_{max}$ in
the c.m. frame \cite{oo97}. The analytic expression can be worked
out as
\ba%
\label{eq:gnem} G(s)=\al\al\frac{1}{16\pi^2s}
\{\sigma(\arctan\frac{s+\Delta}{\sigma\lambda_1}
+ \arctan\frac{s-\Delta}{\sigma\lambda_2}) \noo \\
\al\al- [(s+\Delta)\ln(\frac{q_{max}}{m_1}(1+\lambda_1)) +
(s-\Delta)\ln(\frac{q_{max}}{m_2}(1+\lambda_2))]\}%
\ea%
where $\sigma=[-(s-(m_1+m_2)^2)(s-(m_1-m_2)^2)]^{1/2}$,
$\Delta=m_1^2-m_2^2$ and
$\lambda_i=\sqrt{1+\frac{m_i^2}{q^2_{max}}}$ ($i=1,2$), and the
channel label $ii$ has been dropped.

However, there are some problems. If one goes to values of the
on-shell momenta above the cut-off, one can not use the cut-off
formula \cite{oset}. On the other hand, we find that there is a
severe problem using this cut-off method to calculate the loop
integral. From Eq.~(\ref{eq:gnem}), one can find there is a
singularity above the threshold which is located at
\ba%
\label{eq:sing} (\sqrt{s})_{sing} =
\sqrt{m_1^2+q_{max}^2}+\sqrt{m_2^2+q_{max}^2}.
\ea%
When $\sqrt{s}$ goes to $(\sqrt{s})_{sing}$, the real part of
$G(s)$ will goes to infinity, and when $\sqrt{s}$ is greater than
$(\sqrt{s})_{sing}$ given below in Eq.~(\ref{eq:sing}),
Eq.~(\ref{eq:gnem}) gives a vanishing imaginary part. However, if
momenta are large enough to bring the propagators on shell, the
imaginary part should be given by $-p_{cm}/(8\pi \sqrt{s})$. We
can give some remarks on the singularity. First, after some
manipulations, the real part of the loop integral can be given as
\be%
\text{Re}G(s)=\frac{1}{4\pi^2}{\text P}\int_0^{q_{max}} d|{\bm q}|
\frac{{\bm q}^2(\omega_1+\omega_2)}{\omega_1\omega_2
[s-(\omega_1+\omega_2)^2]},
\ee%
where ${\text P}\int$ represent principal integral, and
$\omega_i=\sqrt{{\bm q}^2+m_i^2}$. One can see that the singularity
comes from the possibility of the denominator in the integrant being
zero. And when the cut-off $q_{max}$ runs to infinity, the
singularity will disappear. Second, the singularity rises from the
$\arctan f(s)$ function. If we perform a series expansion at the
singularity point, the first term should be $\arctan i$, or
equivalently $\frac{1}{2i}\ln 0$ using the identity
$\arctan{z}=\frac{1}{2i}\ln{\frac{z-i}{z+i}}$. It should be a
logarithmic singularity. Third, when $\sqrt{s}=(\sqrt{s})_{sing}$,
the three-momenta of the propagators equal to $q_{max}$. Then no
momentum is available to bring $\sqrt{s}$ larger than
$(\sqrt{s})_{sing}$, and the imaginary part of $G(s)$ should vanish
as it appears in Eq.~(\ref{eq:gnem}). In other words,
$(\sqrt{s})_{sing}$ should be available limit of this method. On the
other hand, there is still a region below $(\sqrt{s})_{sing}$ where
the impact of the singularity is large, and one should be care of
that. For the $S$ wave $K\pi$ scattering, a typical value of the
$q_{max}$ is about 0.6 GeV from fitting to the data when the kernel
of the BSE is of the $O(p^2)$ order. The corresponding position of
the artificial pole will be about 1.4 GeV, and its impact will show
up above 1 GeV apparently. Then it is difficult to describe $K\pi$
data above about 1 GeV in this formalism. Note that this cut-off
method is different from the one used in Pauli-Villars
regularization (see e.g. Ref. \cite{ps}). In this method, the
cut-off is just a three-momentum cut-off.

The way to solve this problem is using a dimensional regularization
method with a dispersion relation to deal with the loop integral.
The analytic expression has been given as \cite{oond,or02}
\ba%
G(s)=\al\al\frac{1}{16\pi^2}\{a(\mu)+\log{\frac{m_1^2}{\mu^2}} +
\frac{\Delta-s}{2s}\log{\frac{m_1^2}{m_2^2}} \noo\\
\al\al +\frac{\sigma}{2s}[\log{(s-\Delta+\sigma)}
+ \log{(s+\Delta+\sigma)} \noo\\
\al\al- \log{(-s+\Delta+\sigma)} - \log{(-s-\Delta+\sigma)}] \},
\ea%
where $a(\mu)$ is a subtraction constant, $\mu$ is the
regularization scale. The result is independent of $\mu$ because
the change causing by a change of $\mu$ can be cancelled by a
change of the subtraction constant $a(\mu)$.

%--------------------------------------------------------------------------------
\begin{figure}[htb]
\begin{center}\vspace*{2.0cm}
{\epsfysize=6cm \epsffile{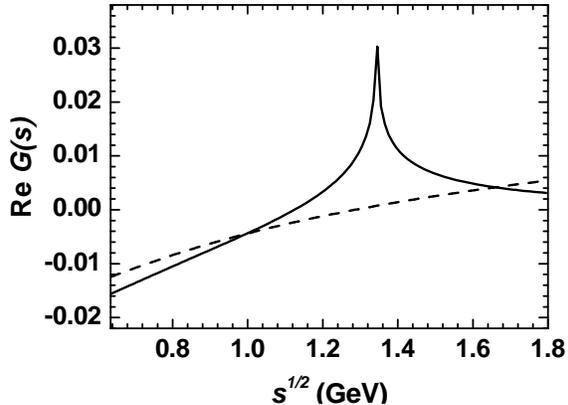}}%
\vglue -3cm\caption{\label{fig1}The real part of the loop integral
calculated using the cut-off method (solid surve) and the
dimensional regularization method (dashed curve).}
\end{center}
\end{figure}
%--------------------------------------------------------------------------------

In order to show the impact of the artificial singularity when the
cut-off method is used, we plot the real part of the loop integral
with $m_1=m_{\pi}$ and $m_2=m_K$ in Fig.~\ref{fig1}. In
Fig.~\ref{fig1}, the solid curve represents the result from the
cut-off method with $q_{max}=0.57$ GeV and the dashed curve
represents the result from the dimensional regularization method
calculated at $\mu=m_K$ with $a(m_k)=-1.41$. From
Eq.~(\ref{eq:sing}), the singularity is located at $\sqrt{s}=1.34$
GeV, and the values above 1 GeV are all affected significantly.
However, the physical results of Ref.~\cite{oo97} where the
cut-off method was used will not be influenced, because with the
parameter $q_{max}=1.03$ GeV therein the singularity will be
located 2.08 GeV, 2.29 GeV and 2.21 GeV for the $\pi\pi$,
$K\bar{K}$ and $\pi\eta$ channels respectively, and the effect
below 1.2 GeV is very small.

\subsection{\label{sec:proof}A simple proof of unitarity}

It has been shown that by this coupled-channel BSE approach
unitarity is ensured \cite{oo99,oond}. In fact, it can be realized
quickly as follows:\\
The unitary relation Eq.~(\ref{eq:ur}) can be rewritten as
\be%
\label{eq:ur2}
\frac{1}{T(s)^*}-\frac{1}{T(s)}=-2i\rho(s).%
\ee%
The general solution of Eq.~(\ref{eq:ur2}) is of the form
\be%
\label{eq:ur3}
T(s)=\frac{1}{K(s)+i\rho(s)},%
\ee%
where $K(s)$ is a real function of $s$. While by the Cutcosky rule
\cite{iz}, the imaginary part of the loop integral can be related
to the phase space factor as
\be%
\text{Im}G(s)=-\rho(s).%
\ee%
So Eq.~(\ref{eq:bse}) can be rewritten as
\ba%
T(s)\al\al=\frac{1}{V^{-1}(s)-G(s)} \noo\\
\al\al=\frac{1}{V^{-1}(s)-\text{Re}G(s)-i\text{Im}G(s)}\noo\\
\al\al=\frac{1}{K(s)+i\rho(s)},%
\ea%
where $K(s)=V^{-1}(s)-\text{Re}G(s)$. In this approach, $V(s)$ is
the lowest order ChPT amplitude and so that is real. From the above
equation one can see in this approach Eq.~(\ref{eq:bse}) ensures
unitarity actually. Furthermore, if one use such a kernel that the
BSE can be dealt with in an on-shell way, the unitary relation can
be ensured if and only if the kernel is real.

\section{\label{sec:res}Results and discussion}

\subsection{\label{sec:ps}$S$ wave $I=1/2$ $K\pi$ phase shifts}

Given the above knowledge, one can calculate the $S$ wave $K\pi$
scattering phase shifts numerically. The only parameter in our
calculations is the subtraction constant $a(\mu)$. In principle,
it should be different for different channels and be same only
under exact $SU(3)$ symmetry. In practice, in order to reduce the
number of free parameters, we take $a(\mu)$ to be same for
different channels. All of the results presented in this paper are
calculated at $\mu=m_K$. In order to get the most accurate
description, we use the MINUIT function minimization and error
analysis package from the CERN Program Library \cite{mint} to find
the most appropriate value of the free parameter $a(m_K)$.

%--------------------------------------------------------------------------------
\begin{figure}[htb]
\begin{center}\vspace*{2.0cm}
{\epsfysize=6cm \epsffile{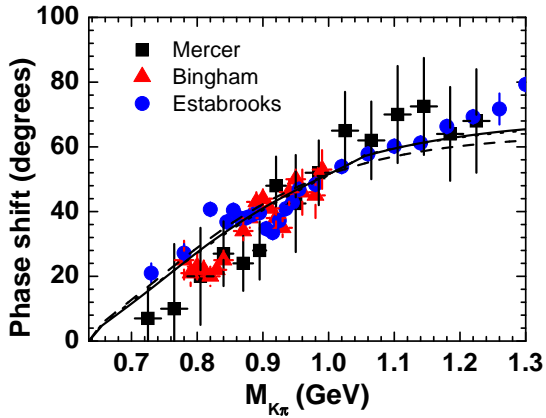}}%
\vglue -2.5cm\caption{\label{fig2}$S$ wave $I=1/2$ $K\pi$ phase
shifts with the dashed line depicting one-channel case, the dotted
line depicting two-channel case, and the solid line depicting
three-channel case, respectively.}
\end{center}
\end{figure}
%--------------------------------------------------------------------------------

The $S$ wave $I=1/2$ $K\pi$ phase shift data are taken from
Refs.~\cite{me71,bi72,es78}. The parameters can be determined from
fitting to the data, and the values are
\ba%
\text{for~one~channel}~~~:~~~~\al\al a(m_K)=-1.412\pm0.017,\noo\\
\text{for~two~channels}~~:~~~~\al\al a(m_K)=-1.278\pm0.014,\noo\\
\text{for~three~channels}:~~~~\al\al a(m_K)=-1.383\pm0.006.
\ea%
The calculated phase shifts are plotted as solid curves in
Fig.~\ref{fig2}, where the dashed line depicts one-channel case,
the dotted line depicts two-channel case, and the solid line
depicts three-channel case, respectively. One can see the data
below $1.2$ GeV can be well described for one channel and two
channels, and the data below $1.2$ GeV can be well described for
three channels. The fit from three-channel calculations is
slightly better.

As mentioned in the previous section, if one use the
three-momentum cut-off method to deal with the loop integral,
there will be an artificial singularity in the loop function. The
typical value of $q_{max}$ is 0.4-0.6 GeV from fitting the phase
shift data below 1 GeV or 1.2 GeV, and the singularity is located
at about 1.0-1.4 GeV. Then the effect of the singularity above 1
GeV is large, see Fig.~\ref{fig1}. That is to say, one cannot get
a satisfactory description of $K\pi$ $S$ wave scattering above 1
GeV using the three-momentum cut-off method. One way out is using
the dimensional regularization to deal with the loop integral as
illustrated above.

\subsection{\label{sec:ps2}$S$ wave $I=3/2$ $K\pi$ phase shifts}

\begin{figure}[htb]
\begin{center}\vspace*{2.0cm}
\epsfysize=6cm \epsffile{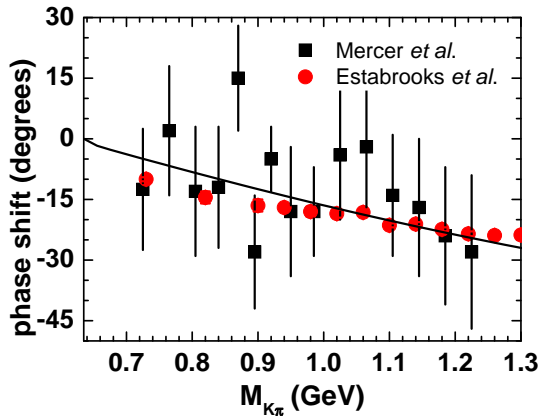}%
\vglue -3cm\caption{\label{fig3}$S$ wave $I=3/2$ $K\pi$ phase
shifts.}
\end{center}
\end{figure}

The experimental data of $S$ wave $I=3/2$ $K\pi$ phase shifts are
taken from Refs.~\cite{me71,es78}. The calculated results are
plotted in Fig.~\ref{fig3}. From fitting to the data, the
subtraction constant is determined to be $a(m_K)=-4.643\pm0.083$.

%--------------------------------------------------------------------------------
%--------------------------------------------------------------------------------

\subsection{Comparison with the results from Roy-Steiner equation analysis}

\begin{figure}[htb]
\begin{center}\vspace*{2.0cm}
\epsfysize=6cm \epsffile{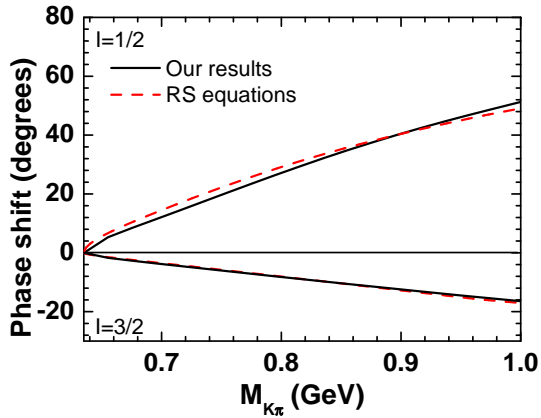}%
\vglue -3cm\caption{\label{fig4}Comparison of $S$ wave $K\pi$ phase
shifts with the results from Roy-Steiner equation analysis. The
upper half and the lower half are for $I=1/2$ and $I=3/2$,
respectively. The solid and dashed lines depict our results and the
RS equation results, respectively.}
\end{center}
\end{figure}

Very recently, some authors have performed a complicated analysis of
$K\pi$ scattering using RS equations \cite{bd04}. The experimental
data above $s=0.935$ GeV$^2$ were used as input, and the solutions
were obtained in the range $s<0.935$ GeV$^2$. The results are very
precise, and it is deserved for our results to be compared with the
RS equation results. For illustration, we plot our results in the
range $s<1$ GeV$^2$  for $I=1/2$ with two channels and for $I=3/2$
as well as the RS equation results \cite{bd04} in Fig. \ref{fig4}.
The upper half and the lower half of Fig. \ref{fig4} are for $I=1/2$
and $I=3/2$, respectively. In this figure, the solid lines represent
the ChUA results, and the dashed lines represent the results from
the detailed analysis using the RS equations \cite{bd04}. From
comparison, we find that our results are consistent with the precise
results in Ref. \cite{bd04} below 1 GeV for both of $I=1/2$ and
$I=3/2$.

%--------------------------------------------------------------------------------
%--------------------------------------------------------------------------------

\subsection{\label{sec:pole}Poles in the amplitudes}

The physical resonances can be associated to the poles in
unphysical Riemann sheets of the scattering amplitudes. After the
same analytic continuation of the loop function $G(s)$ as in
Refs.~\cite{oo97,ro05}, one can find a pole corresponding to the
long controversial scalar particle $\kappa$ which is called
$K^*_0(800)$ in the PDG \cite{pdg04}. The pole position for each
calculation is given in Table~\ref{tab:pole}.

In order to see the stability of the pole position against
different regularization schemes, the results from the cut-off
method are also listed. For each calculation, the parameter
$q_{max}$ is determined from fitting to the phase shift data below
1 GeV where the influence of the artificial singularity is small
as pointed out in the previous section. In the first column of
Table~\ref{tab:pole}, DRM represents the dimensional
regularization method and COM represents the cut-off method. In
the second column of Table~\ref{tab:pole}, the parameter is
$a(m_K)$ for DRM and is $q_{max}$ for COM. The pole positions are
calculated using the central value of each parameter. One can see
that the difference of pole position, especially the real part, is
small using different regularization schemes.

\begin{table}[htb]
\caption{\label{tab:pole} Pole positions found in the second
Riemann sheet of the $T$ matrices.}
\begin{center}
\begin{tabular}{cccc}
\hline\hline
Method & Number of channels & Parameter & Pole position (GeV)\\
\hline
& 1 & $-1.412\pm0.017$ & $0.736-i0.273$ \\
DRM & 2 & $-1.278\pm0.014$ & $0.725-i0.297$ \\
& 3 & $-1.383\pm0.006$ & $0.742-i0.273$ \\
\hline
& 1 & ($0.512\pm0.005$)~GeV & $0.729-i0.244$ \\
COM & 2 & ($0.477\pm0.003$)~GeV & $0.730-i0.244$ \\
& 3 & 0.486~GeV & $0.728-i0.247$ \\
\hline\hline
\end{tabular}
\end{center}
\end{table}

Because in the lowest order, the decay constants are same for
$\pi$, $K$ and $\eta$, one can investigate some effects of higher
order Lagrangian on the pole position by taking the physical
values of the decay constants of $K$ and $\eta$, that is,
$f_K=113.0$ MeV and $f_{\eta}=110.9$ MeV \cite{pdg04}. Let us
study the case of two channels, and the results of the other cases
are similar. The replacement can be done following the Appendix B
of Ref.~\cite{oo99}. For example, $1/f_{\pi}^2$ in
Eq.~(\ref{eq:v11}) will be replaced with $1/(f_Kf_{\pi})$. The
pole positions are listed in Table~\ref{tab:pole2}. The real and
imaginary parts of the pole position are both slightly larger than
the corresponding ones listed in Table~\ref{tab:pole}.

\begin{table}[htb]
\caption{\label{tab:pole2} Pole positions found in the second
Riemann sheet of the two channel $T$ matrices taking different
decay constants for $\pi$, $K$ and $\eta$. }
\begin{center}
\begin{tabular}{ccc}
\hline\hline
Method & Parameter & Pole position (GeV)\\
\hline
DRM & $-1.798\pm0.015$ & $0.790-i0.320$ \\
COM & $0.619\pm0.006$ GeV & $0.773-i0.276$ \\
\hline\hline
\end{tabular}
\end{center}
\end{table}

These results can be compared with previous results in other
chiral unitary approaches. The pole position appeared in the
second Riemann sheet associated to $\kappa$ were found to be
around $770-i250$ MeV in Ref.~\cite{oo99} where COM was used and
$779+i330$ MeV in Ref.~\cite{oond} where DRM was used. These
results are obtained using the $O(p^4)$ chiral amplitudes as the
four-meson contact terms. The position of the pole we found is
similar to the one found in other approaches. Furthermore, from
comparison, it seems that the higher order corrections from the
decay constants give an important contribution to the $O(p^4)$
order corrections to the pole position.

\section{\label{sec:app}Application to the decay $J/\psi\to {\bar K}^{*0}(892)K^+\pi^-$}

\begin{figure}[htb]
\begin{center}\vspace*{3.cm}
\epsfysize=8cm \epsffile{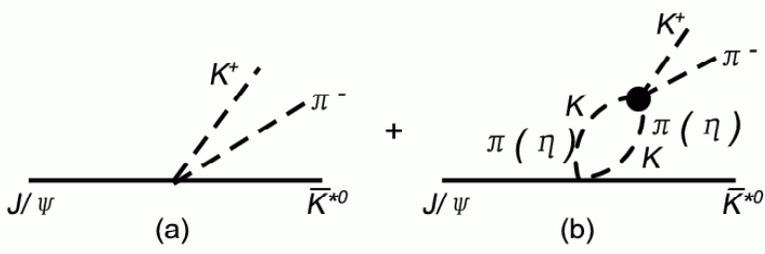}%
\vglue -8cm\caption{\label{fig5}Diagrams for the decay $J/\psi\to
{\bar K}^{*0}(892)K^+\pi^-$.}
\end{center}
\end{figure}

In the preceding sections, we have constructed a formalism of the
$S$ wave $K\pi$ scattering below 1.2 GeV. In this section, as an
example of its applications, it will be applied to describe the $S$
wave $K\pi$ FSI in the decay $J/\psi\to {\bar K}^{*0}(892)K^+\pi^-$.
We will calculate the invariant mass of $K^+\pi^-$ below 1.2 GeV and
fit to the experimental data \cite{bes05}. The relevant decay
mechanisms are plotted in Fig.~\ref{fig5}.

%--------------------------------------------------------------------------------
%--------------------------------------------------------------------------------

In Fig.~\ref{fig5}, (a) represents the direct decay term, and (b)
takes into account the $S$ wave $K\pi$ FSI with coupled channels. We
are interested in the region below 1.2 GeV. In this region, only one
resonance, $K^{*0}(892)$, can decay into $K^+\pi^-$. However, the
branching ratio of $J/\psi\to K^{*0}(892){\bar K}^{*0}(892)$ is very
small. As a simple consideration, no intermediate resonances will be
considered.

Similar to Refs.~\cite{mo01,rp04}, the contact terms of the vertex
$J/\psi VPP$ can be extracted from a Lagrangian of the form
\be%
\label{eq:lc} {\cal L}_c = g\psi_{\mu}\langle
V^{\mu}\Phi\Phi\rangle,
\ee%
where $\psi_{\mu}$, $V^{\mu}$ and $\Phi$ are the fields of
$J/\psi$, vector and pseudoscalar octets respectively. $V$ and
$\Phi$ are both $3\times3$ matrices with
\be%
\label{eq:V} V = \left(
\begin{array}{ccc}
\frac{1}{\sqrt{2}}\rho^0 + \frac{1}{\sqrt{2}}\omega & \rho^+ & K^{*+}\\
\rho^- & -\frac{1}{\sqrt{2}}\rho^0 + \frac{1}{\sqrt{2}}\omega & K^{*0}\\
K^{*-} & \bar{K}^{*0} & \phi
\end{array}
\right),
\ee%
where ideal mixing between $\phi$ and $\omega$ is assumed, and
$\Phi$ has been given in Eq.~(\ref{eq:ps}). The relevant terms in
Eq.~(\ref{eq:lc}) can be written as
\be%
\label{eq:direct} g\psi_{\mu}{\bar K}^{*0\mu}(K^+\pi^- -
\frac{1}{\sqrt{2}}K^0\pi^0 - \frac{1}{\sqrt{6}}K^0\eta).
\ee%
Let us denote the momenta of $J/\psi$, $K^+$, $\pi^-$ and ${\bar
K}^{*0}(892)$ with $p,~p_1,~p_2$ and $p_3$, respectively. From
Eq.~(\ref{eq:direct}), the amplitude for Fig.~\ref{fig5}(a) is of
$S$ wave, and it is
\be%
\label{eq:cont} t_0 = -g
\varepsilon_{\mu}^{(\lambda)}(p)\varepsilon^{\mu(\lambda')}(p_3),
\ee%
where $\varepsilon_{\mu}^{(\lambda)}(p)$ and
$\varepsilon^{\mu(\lambda')}(p_3)$ are the polarization vectors of
$J/\psi$ and ${\bar K}^{*0}$.

From the preceding sections, the $S$ wave $I=1/2$ $K\pi$ phase shift
data below 1.2 GeV can be fitted well using two coupled channels.
The coupled channel $S$ wave $K\pi$ FSI can be taken into account
using the formalism. After that, the amplitude corresponding to
Fig.~\ref{fig5}(a) and (b) is
\ba%
\label{eq:fsi}
t_c\equiv\al\al
-g\varepsilon_{\mu}^{(\lambda)}(p)\varepsilon^{\mu(\lambda')}(p_3)
F(s_{12})\noo\\
=\al\al-g\varepsilon_{\mu}^{(\lambda)}(p)\varepsilon^{\mu(\lambda')}(p_3)
(1 + G_{11}\langle K^+\pi^-|t|K^+\pi^-\rangle \noo\\
\al\al - \frac{1}{\sqrt{2}} G_{11}\langle
K^+\pi^-|t|K^0\pi^0\rangle - \frac{1}{\sqrt{6}} G_{22}\langle
K^+\pi^-|t|K^0\eta\rangle),
\ea%
where $\langle K^+\pi^-|t|P_1P_2\rangle$ ($P_1P_2=K^+\pi^-$, etc.)
represents the full amplitude of the process $P_1P_2\to K^+\pi^-$
calculated in the ChUT, and $s_{12}=(p_1+p_2)^2=m_{12}^2$. Because
the isospin of $J/\psi$ is 0 and of ${\bar K}^{*0}$ is 1/2, the
FSI amplitudes must be of isospin 1/2. These amplitudes can be
related to the amplitudes in the isospin basis as
\ba%
\al\al\langle K^+\pi^-|t|K^+\pi^-\rangle^{I=1/2}=\frac{2}{3}T_{11}^{1/2}(s),\\
\al\al\langle K^+\pi^-|t|K^0\pi^0\rangle^{I=1/2}=-\frac{\sqrt{2}}{3}T_{11}^{1/2}(s),\\
\al\al\langle
K^+\pi^-|t|K^0\eta\rangle^{I=1/2}=-\sqrt{\frac{2}{3}}T_{12}^{1/2}(s),
\ea%
where $T_{ij}^{(1/2)}(s)$ is the full amplitude of the channel
$j\to i$ ($i=1: K\pi$, $i=2: K\eta$).

The invariant mass spectrum of $K^+\pi^-$ can be calculated using
the formula \cite{pdg04}
\be%
\frac{d\Gamma}{dm_{12}}=\frac{1}{(2\pi)^5}\frac{1}{16M^2}
\overline{\sum_{\lambda}}\sum_{\lambda'}\int |t_c|^2|{\bf
p_1^*}||{\bf p_3}|d\Omega^*_1d\Omega_3,
\ee%
where $\overline{\sum}_{\lambda}\sum_{\lambda'}$ describes the
average over initial states and the sum over final states, $(|{\bf
p_1^*}|, \Omega^*_1)$ is the momentum of $K^+$ in the c.m. frame
of $K^+\pi^-$, and $(|{\bf p_3}|, \Omega_3)$ is the momentum of
${\bar K}^{*0}(892)$ in the rest frame of $J/\psi$.

On the other hand, because the experimental data are without
sideband subtraction \cite{bes05}, we will consider the sequential
process $J/\psi\to K^{*0}(892)K^-\pi^+\to K^+\pi^-K^-\pi^+$ as a
sideband background.

For the process $J/\psi\to K^{*0}K^-\pi^+$, one can write an
effective amplitude like the one in Eq.~(\ref{eq:fsi}).
\ba%
t_c = -g\varepsilon_{\mu}^{(\lambda)}(p)
\varepsilon^{\mu(\lambda')}(p_K^{*0}) F(s_{34}),
\ea%
where $s_{34}=(p_{K^-}+p_{\pi^+})^2=m_{34}^2$.

The amplitude of the $P$ wave decay $K^{*0}(892)\to K^+\pi^-$ can
be written as
\be%
t_{K^{*0}K^+\pi^-}=g_{K^{*}K\pi}\varepsilon^{(\lambda'')}(p_4)\cdot(p_2-p_1).
\ee%
From the width of $K^{*0}(892)$, the dimensionless coupling
constant can be determined to be $g_{K^{*}K\pi}=4.61$. Now one can
write the amplitude for the sideband background process as
\ba%
t_s = -gg_{K^{*}K\pi}F(s_{34})\varepsilon_{\mu}^{(\lambda)}(p)
\frac{-g^{\mu\nu}+(p_1+p_2)^{\mu}(p_1+p_2)^{\nu}/s_{12}}
{s_{12}-m_{K^*}^2+i\sqrt{s_{12}}\Gamma_{K^*}(s_{12})}
(p_2-p_1)_{\nu},
\ea%
where $\Gamma_{K^*}(s_{12})$ is the momentum-dependent width of
$K^{*0}(892)$. The relation of $\Gamma_{K^*}(s_{12})$ and the
on-shell width $\Gamma_{K^*}(m_{K^*}^2)=50.7$ MeV is
\be%
\Gamma_{K^*}(s_{12})=\Gamma_{K^*}(m_{K^*}^2)
\frac{m_{K^*}^2}{s_{12}} \frac{q^3}{q_0^3},
\ee%
where $q$ and $q_0$ is momentum of one meson in the c.m. frame
\ba%
\al\al\label{eq:qcm}
q=\frac{1}{2\sqrt{s_{12}}}\sqrt{(s_{12}-(m_K+m_{\pi})^2)(s_{12}-(m_K-m_{\pi})^2)},\\
\al\al
q_0=\frac{1}{2m_{K^*}}\sqrt{(m_{K^*}^2-(m_K+m_{\pi})^2)(m_{K^*}^2-(m_K-m_{\pi})^2)}.
\ea%

Since $K^+$ and $\pi^-$ in their c.m. frame are in $S$ wave in the
direct decay process and in $P$ wave in the sideband background
process, there is no interference between the amplitudes $t_c$ and
$t_b$. Hence the invariant mass of $K^+\pi^-$ can be calculated
using the formula \cite{pdg04}
\ba%
\label{eq:wid} \frac{d\Gamma}{dm_{K\pi}} =
\al\al\frac{1}{(2\pi)^5}\frac{1}{16M^2}
\overline{\sum_{\lambda}}\sum_{\lambda'}\int |t_c|^2 |{\bf
p_1^*}||{\bf p_3}|d\Omega^*_1d\Omega_3 \noo\\
\al\al + \frac{1}{(2\pi)^8}\frac{1}{32M^2}\int\int^{b}_{a}dm_{34}
\overline{\sum_{\lambda}} |t_s|^2|{\bf p_1^*}||{\bf
p_3^{'*}}|d\Omega^*_1d\Omega_3^{'*}d\Omega_{12}
\ea%
where $\overline{\sum}_{\lambda}\sum_{\lambda'}$ describes the
average over initial states and the sum over final states, $(|{\bf
p_1^*}|, \Omega^*_1)$ is the momentum of $K^+$ in the c.m. frame
of $K^+\pi^-$, $(|{\bf p_3}|, \Omega_3)$ is the momentum of ${\bar
K}^{*0}(892)$ in the rest frame of $J/\psi$, and $(|{\bf
p_3^{'*}}|, \Omega_3^{'*})$ is the momentum of $K^-$ in the c.m.
frame of $K^-\pi^+$. In the integral $\int^{b}_{a}dm_{34}$ in
Eq.~(\ref{eq:wid}), $a=0.812$ GeV and $b=0.972$ GeV according to
the experimental events selection criterion 0.812 GeV
$<m_{K^-\pi^+}<$ 0.972 GeV \cite{bes05}.

%--------------------------------------------------------------------------------
\begin{figure}[htb]
\begin{center}\vspace*{2.0cm}
\epsfysize=6cm \epsffile{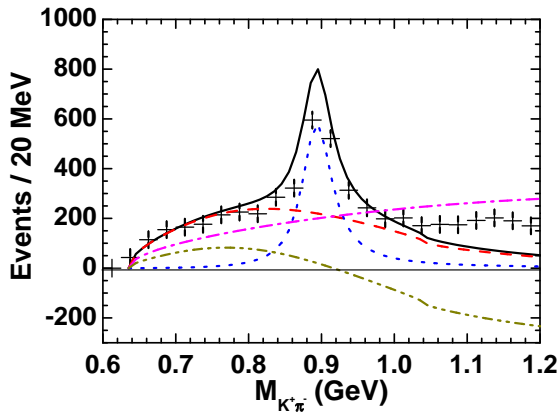}%
\vglue -3cm\caption{\label{fig6}The invariant mass spectrum of
$K^+\pi^-$ below 1.2 GeV of the decay $J/\psi\to {\bar
K}^{*0}(892)K^+\pi^-$. The black dots with error bars are the
experimental data. The solid line represents the fit result, the
dash-dotted, dashed and dotted lines represent the contributions of
the direct term, the direct term $+$ the $S$ wave $K\pi$ FSI and the
sideband background process $J/\psi\to K^{*0}(892)K^-\pi^+\to
K^+\pi^-K^-\pi^+$, respectively. The dash-dot-dotted line represents
the total contributions of the FSI and the interference of the FSI
and the direct term.}
\end{center}
\end{figure}
%--------------------------------------------------------------------------------

There is only one parameter $g$. The program package MINUIT
\cite{mint} is used to make a fit and find the most appropriate
value of the parameter. The BES data are taken from Ref.
\cite{bes05}. The results are plotted in Fig.~\ref{fig6} and the fit
is good qualitatively. The solid line represents the fit result, the
dash-dotted and dashed lines represent the contributions of the
direct term and the direct term $+$ the $S$ wave $K\pi$ FSI,
respectively. The dotted line represents the background or sideband
contribution coming from $J/\psi\to K^{*0}K^-\pi^+\to
K^+\pi^-K^-\pi^+ $ in the region 0.812 GeV $\leq
m_{K^-\pi^+}\leq0.972$ GeV. The dash-dot-dotted line represents the
contributions concerning $\kappa$ or of the FSI and the interference
of the FSI and the direct term. One can see the wide bump below 1.1
GeV besides the peak of $K^{*0}(892)$ is due to $\kappa$.

A decay amplitude can be expressed as a sum of a background and some
Breit-Wigner amplitudes associated to the relevant resonances. From
the experimental data plotted in Fig.~\ref{fig6}, one can see that
there is a wide bump below 1.1 GeV besides the peak of
$K^{*0}(892)$. Were there no $\kappa$ and were the background of no
structure and just like a form of Eq.~(\ref{eq:cont}), this pump
would be hardly understood. From our results plotted in
Fig.~\ref{fig6}, the wide bump is generated by the $S$ wave $K\pi$
FSI. Note that in our ChUT approach, the scalar particle $\kappa$ is
generated dynamically by the $K\pi$ FSI. That is to see, the bump is
generated by $\kappa$.

The value of the parameter determined from fitting to the spectrum
of the number of events is $g=1770.3$. However, it is not
physical, and so far we cannot determine its physical value
because the branching ratio $B(J/\psi\to {\bar
K}^{*0}(892)K^+\pi^-)$ has not been known yet.

\section{\label{sec:con}Conclusions}

We have used a chiral unitary approach to study $S$ wave $K\pi$
scattering and gotten a satisfactory description of phase shift
data below 1.2 GeV. On-shell BSE is used, and the lowest order
ChPT amplitudes are used as its kernel. One-channel, two-channel
and three-channel calculations are performed for the $I=1/2$
$K\pi$ scattering, and one-channel for $I=3/2$. There is only one
parameter $a(m_K)$ for each isospin in the model, and its value is
determined from fitting to the data. A pole of the $I=1/2$ $T$
matrix is found in the second Riemann sheet. We investigate the
pole position with different number of channels with both cut-off
and dimensional regularization schemes, and find that the pole
position is stable against different regularization schemes. If we
take into account some higher order corrections, the real and
imaginary parts of the pole are both slightly enhanced. This pole
corresponds to the long controversial broad resonance $\kappa$. In
this way, the mass and width of $\kappa$ are estimated to be about
$757\pm33$ MeV and $558\pm82$ MeV. It is pointed out that there
would be a singularity in the loop function if the conventional
cut-off method were used. The singularity comes from that the
three-momentum cut-off set a limit for the applicable region.

One advantage of this approach is that there is only one parameter.
Once this parameter is determined from the phase shift data, it can
be used to other processes such as heavy particle decays involving
$S$ wave $K\pi$ or $K\eta$ FSI. In recent years, there are other
excellent works on $K\pi$ scattering, such as \cite{jop00,bd04}. In
Ref.~\cite{jop00}, the resonance chiral Lagrangian together with a
unitarization \cite{oond} was used, and the data of $S$ wave $K\pi$
scattering below 2 GeV can be fitted well. However, the number of
free parameters is large. The pole position corresponding to the
$\kappa$ is about $700-i300$ MeV. It is similar to the one found in
our work. The difference of the mass is large slightly. However, the
width of the resonance is about 600 MeV, and this makes a small
difference of the mass and width is of little importance. In
Ref.~\cite{bd04}, the authors solved a set of six RS equations for
the $S$ and $P$ waves of the $K\pi$ scattering amplitudes. The input
was high-statistical data at $\sqrt{s}\gtrsim1$ GeV for both $\pi
K\to\pi K$ and $\pi\pi\to K{\bar K}$ amplitudes, then the amplitudes
near threshold were determined. Certainly, this is an important
work, and we compared our results for $S$ wave $K\pi$ scattering
phase shifts below 1 GeV with the results in this work. From
comparison, we find that our results are consistent with the precise
results in Ref. \cite{bd04}. Furthermore, the one-parameter model
can be used easily to deal with $S$ wave $K\pi$ FSI appeared in
other processes below $m_{K\pi}=1.2$ GeV. As an example, we
calculated the invariant mass spectrum of $K^+\pi^-$ of the decay
$J/\psi\to {\bar K}^{*0}(892)K^+\pi^-$, and the fit to the
experimental data is qualitatively good below 1.1 GeV. From the
analysis of different contributions from different processes, we
found that $\kappa$ is important to produce the wide bump below 1.1
GeV.

\begin{acknowledgments}
We are very grateful to valuable discussions with E. Oset. We
acknowledge B. Moussallam for providing their Roy-Steiner equation
analyzed data in Ref. \cite{bd04}. This project is partially
supported by the NSFC grant Nos. 90103020, 10475089, 10435080,
10447130 and CAS Knowledge Innovation Key-Project grant No.
KJCX2SWN02.
\end{acknowledgments}

\end{document}